\documentclass[11pt,a4paper]{article}
\usepackage[english]{babel}
\usepackage[mathscr]{eucal}
\usepackage{amsmath,amsbsy,amsthm,amsfonts,amssymb,makeidx,enumerate,bm,wasysym}
\usepackage{courier,vmargin,graphicx,wrapfig,float}
\usepackage{caption,subcaption}
\usepackage{epstopdf}

\numberwithin{equation}{section}

\setmarginsrb{2.5 cm}{1.3 cm}
             {2.5 cm}{1.5 cm}
             {1.2 cm}{0.8 cm}
             {1.2 cm}{0.8 cm}

\def\t{t}
\def\ww{w}
\def\x{\textbf{x}}
\def\y{\textbf{y}}
\def\b0{\mathbf{0}}
\def\de{\partial}
\def\lap{\Delta}
\def\BB{\mathcal{B}}
\def\G{\mathit{G}}
\def\at{\lambda}
\def\ee{\varepsilon}
\def\Dom{\mbox{Dom}}
\def\AA{\mathcal{A}}
\def\AAe{\AA_{\ee}}
\def\u{u}
\def\mm{\kappa}
\def\ga{\gamma}
\def\Om{\Omega}
\def\vfi{\varphi}
\def\Ga{\Gamma}
\def\R{\mathbb{R}}
\def\C{\mathbb{C}}
\def\Fi{\hat{\phi}}
\def\Fiue{\Fi^{\u}_{\epsilon}}
\def\Ti{\hat{T}}
\def\Tiue{\Ti^{\u,\ee}}
\def\rist{\upharpoonright}
\def\geqs{\geqslant}
\def\l{\left}
\def\r{\right}
\def\la{\langle}
\def\ra{\rangle}
\def\dd{\displaystyle}
\def\Im{\mbox{Im}\,}
\def\Re{\mbox{Re}\,}

\title{\textbf{THE CASIMIR ENERGY ANOMALY\\ FOR A POINT INTERACTION}\vspace{0.5cm}}

\author{
{\Large Davide Fermi}\vspace{0.1cm} \\
Dipartimento di Matematica, Universit\`a degli Studi di Milano\\
Via Saldini 50, I-20133 Milano, Italy\\
and\\
Istituto Nazionale di Fisica Nucleare, Sezione di Milano,\\
Via Celoria 16, I-20133 Milano, Italy \vspace{0.2cm}\\
e--mail: davide.fermi@unimi.it\vspace{0.3cm}
}

\date{}

\begin{document}
\maketitle

\begin{abstract} 
The Casimir energy for a massless, neutral scalar field in presence of
a point interaction is analyzed using a general zeta-regularization
approach developed in earlier works. In addition to a regular bulk contribution,
there arises an anomalous boundary term which is infinite despite renormalization.
The intrinsic nature of this anomaly is briefly discussed.
\end{abstract}
\vspace{0.3cm} \noindent
\vspace{0.15cm}\\
\textbf{MSC\! 2010}: 81T55; 81T10; 81Q10.
\vspace{0.15cm}\\
\textbf{PACS\! 2010}: 03.70.+k; 11.10.Gh; 03.65.Db.
\vspace{0.9cm}

\section{Introduction}

This note deals with the Casimir effect for a massless, neutral scalar field in presence of an external delta-type potential concentrated at a point. Similar models were previously considered in the literature \cite{Alb2,Alb1,BorPir,FPDelta,Grats1,Grats2,Scar,SprZer}, building on various mathematically sound descriptions of the Schr\"odinger operator comprising the said singular potential (see, e.g., \cite{AlbB,BeFa,BM2015}).

Continuing the analysis begun in Ref.~\cite{FPDelta}, here the total Casimir energy for the above model is investigated within a general framework for zeta regularization developed in previous works \cite{CFP,FP1,FP2,FP3,FPBook}. In addition to a regular bulk contribution which is finite after renormalization, there also appears an anomalous boundary term which remains infinite even after implementing the standard renormalization procedure. The arising of this anomaly is ascribed to an unnatural interpretation of the model.
\vspace{0.1cm}

\section{The Reference Model and Zeta Regularization}

Consider $(1\!+\!3)$-dimensional Minkowski spacetime, endowed with a set of inertial coordinates $(x^\mu)_{\mu = 0,1,2,3} \!=\! (t, \x) \in \R\!\times\!\R^3$, such that the metric has components $(\eta_{\mu\nu})= \mbox{diag}(-1, 1, 1, 1)$ (natural units are employed, so that $c \!=\! 1$ and $\hbar \!=\! 1$).
\eject\vfill\noindent

The theory of a canonically quantized, massless, neutral scalar field $\Fi$ living on Minkowski spacetime in presence of an external delta-type potential concentrated at the point $\x = \b0$ can be described making reference to the space domain \mbox{$\Om = \R^3 \backslash \{\b0\}$} and considering the Klein-Gordon equation $(-\,\de_{tt} + \AA)\,\Fi \!=\! 0$, where $\AA$ is a \mbox{self-adjoint} realization on $L^2(\Om)$ of the $3$-dimensional Laplacian $-\lap$, accounting for suitable boundary conditions at $\x = \b0$. More precisely, one has (see \cite{AlbB,FPDelta})
\begin{equation}
\begin{array}{c}
\dd{\Dom\,\AA := \Big\{\psi = \vfi + {4 \pi \at \over 1 -  i\sqrt{z}\,\at}\, \vfi(\b0)\, \G_z
\,\Big|\,z \!\in\! \C \backslash [0,\infty),\, \vfi \!\in\! H^2(\R^3)\Big\} \,,} \vspace{0.1cm} \\
\dd{\AA := (-\lap) \!\rist\! \Dom\AA \,\subset L^2(\Om) \to L^2(\Om) \,,}
\end{array} \label{defAAa}
\end{equation}
where $\at$ is a real parameter related to the strength of the potential, $\G_z \!:=\! {e^{i \sqrt{z}\,|x|} \over 4\pi |\x|}$ with $\Im\! \sqrt{z} \!>\! 0$ and $H^2(\R^3)$ is the usual Sobolev space of order two. The non-negativity of $\AA$ (necessary for a consistent formulation of the field theory; see \cite{FPBook}) is ensured assuming $\at \!\geqs\! 0$; in this case, $\AA$ has purely absolutely continuous spectrum $\sigma(\AA) \!=\! [0,\infty)$. Besides, note that the choice $\at \!=\! 0$ corresponds to the free theory where no delta potential is present (correspondingly, $\Dom\AA|_{\at = 0} \!= \!H^2(\R^3)$).

Next, consider the modified operator $\AAe \!:=\! \AA + \ee^2$, where the fictitious mass $\ee \!>\! 0$ plays the role of an infrared cut-off. For $\t \!>\! 0$, the associated heat kernel reads (see \cite{AlbHeat, FPDelta})
\begin{equation}
\begin{array}{c}
\dd{e^{-\t\AAe}(\x,\y)} \vspace{0.05cm}\\
\dd{= {e^{-\,\ee^2 \t} \over (4\pi \t)^{3/2}} \l[
e^{-{|\x-\y|^2 \over 4\t}}
\! + {2\,\t \over |\x|\,|\y|}\!\l(\!e^{-{(|\x|+|\y|)^2 \over 4\t}}
\! - {1 \over \at}\! \int_0^{\infty}\!\!\! d\ww\;e^{-\l(\!{\ww \over \at}+{(\ww + |\x|+|\y|)^2 \over 4\t}\!\r)}\!\r)\r].}
\end{array} \label{heatexp}
\end{equation}
Note that the term multiplying ${2\,\t \over |\x|\,|\y|}$ in the above expression vanishes for $\at = 0$.

Replacing $\Fi$ with the zeta-regularized field $\Fiue \!:=\! (\mm^{-2} \AA_{\ee})^{-\u/4}\, \Fi$ ($\u \!\in\! \C$ is the regulating parameter, $\mm \!>\! 0$ is a mass scale parameter; see \cite{FPBook}), one obtains regularized observables whose vacuum expectation values (VEVs) can be expressed in terms of (derivatives of) the integral kernels $\AAe^{-(1\pm \u)/2}(\x,\y)$, evaluated at $\y \!=\! \x$ for $\Re \u$ large enough. Renormalization of these VEVs is attained computing the regular part of their analytic continuation w.r.t.~$\u$ at $\u \!=\! 0$ (\footnote{By definition, the regular part at $\u \!=\! 0$ of a given meromorphic function $f$ with Laurent expansion $f(\u) = \sum_{n = -1}^{\infty}\!f_n\, \u^n$ is $RP|_{\u = 0} f(\u) \!:=\! f_0$.}) and then taking the limit $\ee \!\to\! 0^+$ (see \cite{FPBook}).

The above strategy was employed in Ref.~\cite{FPDelta} to determine the renormalized VEV of the stress-energy tensor. Notably, the renormalized energy density $\la 0 | \Ti_{00}(\x)| 0 \ra_{ren}$ was shown to diverge in a non-integrable way near $\x \!=\! \b0$; so, integrating it over $\Om \!=\! \R^3 \backslash\{\b0\}$ yields an infinite total energy. An alternative approach to treat this quantity entails integrating first the regularized density $\la 0 | \Tiue_{00}(\x)| 0 \ra$ and then addressing renormalization. In this approach, the regularized total energy $\int_{\Om} d\x\,\la 0 | \Tiue_{00}(\x)| 0 \ra$ is given by the sum of bulk and boundary contributions, respectively defined as (see \cite{FPBook})
\begin{gather}
E^{\u,\ee}\! := {\mm^\u\! \over 2}\! \int_{\Om} d\x\; \AAe^{-{\u - 1 \over 2}}\!(\x,\y)\big|_{\y = \x}\,, \label{Eu} \\
B^{\u,\ee}\! := \Big({1 \over 4}-\xi\Big)\, \mm^\u\!\! \int_{\de \Om}\!\! d\sigma(\x)\;\de_{n_\y}\AAe^{-{\u + 1 \over 2}}\!(\x,\y)\big|_{\y = \x} \label{Bu}
\end{gather}
($\xi$ is the conformal parameter, $\de\Om$ is the boundary of $\Om$, $d\sigma(\x)$ is the induced measure on $\de\Om$ and $n_\y$ is the outer unit vector normal to $\de\Om$ at $\y$). 
Finally, recall the Mellin-type identities holding for $\Re \u$ large ($\Ga$ is the Euler's gamma function) (see \cite{FPBook})
\begin{equation}
\AAe^{-{\u \pm 1 \over 2}}\!(\x,\y) = {1 \over \Ga({\u \pm 1 \over 2})} \int_0^{\infty}\!\!\!d\t\;\t^{{\u \pm 1 \over 2} - 1}\,e^{-\t\AAe}(\x,\y)\,.\label{Mel} \vspace{0.4cm}
\end{equation}

\section{The Relative Bulk Energy}\label{sec:Eu}

The ``bare'' regularized bulk energy $E^{\u,\ee}$ diverges for all $\u \!\in\! \C$ due to a potential-independent empty space contribution, present even for $\at \!=\! 0$. Subtracting the latter, one can consider the regularized ``relative'' bulk energy (cf. Eq.~\eqref{Eu})
\begin{equation}
\Delta E^{\u,\ee}\! := {\mm^\u\! \over 2}\! \int_{\Om} d\x\; \Big(\AAe^{-{\u - 1 \over 2}}\!(\x,\x) - \AAe^{-{\u - 1 \over 2}}\!(\x,\x)\big|_{\at = \b0}\Big)\,. \label{DEu}
\end{equation}
From Eqs.~\eqref{heatexp}, \eqref{Mel} and \eqref{DEu}, passing to spherical coordinates (with $r \!:=\! |\x|$) one gets
\begin{equation}
\Delta E^{\u,\ee} \!= \!
{\mm^\u\! \over \sqrt{4\pi}\, \Ga({\u - 1 \over 2})}\! \int_{0}^{\infty}\!\!\!\! dr\! \int_0^{\infty}\!\!\!\! d\t\;\t^{{\u \over 2} - 2} e^{- \ee^2 \t} \!\l(\!e^{-{r^2 \over \t}}
\! - {1 \over \at}\! \int_0^{\infty}\!\!\!\! d\ww\;e^{-\l(\!{\ww \over \at}+{(\ww + 2r)^2 \over 4\t}\!\r)}\!\r). 
\end{equation}
The above representation makes sense for $\Re \u \!>\! 1$; besides, the order of integration can be permuted arbitrarily by Fubini's theorem. Evaluating explicitly the integrals in $r$ and $w$, posing $\tau \!:=\! \sqrt{t}/\at$ and integrating by parts twice one infers
\begin{equation}
\Delta E^{\u,\ee} = 
{(\at\,\mm)^{\u} \over 4\at\,\u\, \Ga({\u + 1 \over 2})} \int_0^{\infty}\!\!\! d \tau\; \tau^{\u}\,
{d^2 \over d\tau^2}\Big(e^{- (\ee \at)^2 \tau^2} e^{\tau^2} \mbox{erfc}(\tau)\Big)\,. 
\end{equation}
In view of the regularity and asymptotic features of the complementary error function $\mbox{erfc}(\tau)$, the latter expression provides the analytic continuation of $\Delta E^{\u,\ee}$ to a meromorphic function of $\u$ for $\Re \u \!>\! -1$, with a simple pole at $\u \!=\! 0$. Taking the regular part and evaluating the limit $\ee \!\to\! 0^+$ (following the general approach of Ref.~\cite{FPBook}), by dominated convergence theorem one obtains the renormalized relative bulk energy
\begin{equation}\begin{array}{rl}
\dd{\Delta E^{ren} :=}\! & \!\!\dd{\lim_{\ee \to 0^+} RP\big|_{\u = 0} \,\Delta E^{\u,\ee} } \vspace{0.05cm} \\
\dd{=}\! & \!\!\dd{{1 \over \at}\! \int_0^{\infty}\!\!\! d \tau\; {\ga\! +\! 2 \log(2 \mm \at \tau) \over 8\sqrt{\pi}} \,
{d^2 \over d\tau^2}\Big(e^{\tau^2}\! \mbox{erfc}(\tau)\Big) = {\log(\mm \at) \over 2\pi \at}}
\end{array}
\end{equation}
($\ga$ is the Euler-Mascheroni constant and the integral was evaluated using \verb"Mathematica"). 


The ``vacuum energy'' $E_{vacuum} \!:=\! - \lim_{\beta \to \infty} \de_{\beta}\! \log Z \!=\! 2\alpha(1\!-\!\log(4\pi\alpha \ell)$ determined in accordance with Ref.~\cite{SprZer} coincides with $\Delta E^{ren}$ if the renormalization length scale of \cite{SprZer} is fixed as $\ell = e/\mm$ (recall also that $\alpha \!=\! 1/(4\pi\at)$, according to Ref.~\cite{FPDelta}).

\section{The Anomalous Boundary Energy Term}\label{sec:Bu}

The space domain $\Om \!=\! \R^3 \backslash\{\b0\}$ has improper boundaries at $|\x| \!\to\! \infty$ and $|\x| \!\to\! 0$. Taking this and spherical symmetry into account, one can express the regularized boundary energy of Eq.~\eqref{Bu} via appropriate limits of integrals over finite-size spheres:
\begin{gather}
B^{\u,\ee} = \Big({1 \over 4}-\xi\Big) \Big[ \lim_{r \to \infty} \BB_{out}^{\u,\ee}(r) + \lim_{r \to 0}\, \BB_{in}^{\u,\ee}(r)\Big]\,; \label{Bur0}\\
\BB_{out/in}^{\u,\ee}(r) \,:=\, \mm^\u\! \int_{\{|\x| \,=\, r\}}\hspace{-0.5cm} d\sigma(\x)\;\de^{out/in}_{n_\y}\AAe^{-{\u + 1 \over 2}}\!(\x,\y)\big|_{\y = \x} \;\quad (r \!>\! 0)\,. \label{Bur}
\end{gather}
In Eq.~\eqref{Bur}, $\de^{out}_{n_\y}$ (resp. $\de^{in}_{n_\y}$) denotes the derivative in the outer (resp. inner) radial direction normal to the sphere $\{|\x| \!=\! r\}$. From Eqs.~\eqref{heatexp}, \eqref{Mel} and \eqref{Bur} one infers
\begin{equation}
\begin{array}{rl}
\dd{\BB_{out/in}^{\u,\ee}(r) =} & \dd{(-/+)\,{1 \over r}\,{\mm^\u \over \sqrt{\pi}\, \Ga({\u + 1 \over 2})} \int_0^{\infty}\!\!\!d\t\;\t^{{\u \over 2} - 1}\, e^{-\,\ee^2 \t}} \\
& \;\; \dd{\times\! \left[\!\l(\!1 \!+\! {r^2 \over t}\r)\! e^{-{r^2\! \over \t}}
\!-\! {1 \over \at} \! \int_0^{\infty}\!\!\!\! d\ww\;e^{-\l(\!{\ww \over \at}+{(\ww + 2r)^2 \over 4\t}\!\r)}\! \l(\!1 \!+\! {r\,(\ww\!+\!2r) \over 2\t}\!\r)\! \right] .}
\end{array} \label{Burexp}\vspace{0.2cm}
\end{equation}
Notably, $\BB_{out/in}^{\u,\ee}(r)  \equiv 0$ for $\at = 0$; thus, there is no empty space contribution to $B^{u,\ee}$.

On one hand, one easily infers by dominated convergence that $\lim_{r \to \infty}\! \BB_{out}^{\u,\ee}(r) \!=\! 0$ for any $\u \!\in\! \C$ (and $\ee \!>\! 0$). On the other hand, again by dominated convergence (though more careful estimates are demanded here), for any $\Re \u \!>\! 0$ one gets
\begin{equation}
\begin{array}{rl}
\dd{\lim_{r \to 0^+}\!\! \big(r\, \BB_{in}^{\u,\ee}(r)\big) =} & \dd{{\mm^\u \over \sqrt{\pi}\, \Ga({\u + 1 \over 2})} \int_0^{\infty}\!\!\!d\t\;\t^{{\u \over 2} - 1}\, e^{-\,\ee^2 \t} \left[1\!-\! {1 \over \at} \! \int_0^{\infty}\!\!\!\! d\ww\;e^{-\l(\!{\ww \over \at}+{\ww^2 \over 4\t}\!\r)} \right] .}
\end{array} \label{Burexplim0}
\end{equation}
Since the r.h.s.\! is finite and not zero (in fact, it involves the integral of a positive function), Eq.~\eqref{Burexplim0} implies $\lim_{r \to 0^+}\! \BB_{in}^{\u,\ee}(r) \!=\! \infty$ for any $\u \!\in\! \C$. This divergence entails an infinite contribution to the renormalized boundary energy; namely,
\begin{equation}
B^{ren} := \lim_{\ee \to 0^+} RP\big|_{\u = 0} B^{\u,\ee} = \lim_{\ee \to 0^+} RP\big|_{\u = 0} \Big(\lim_{r \to 0^+}\! \BB_{in}^{\u,\ee}(r)\Big) = \infty\,.
\end{equation}

Thus, while no contribution arises from spatial infinity (as was to be expected), an infinite energy occurs where the potential is concentrated. Conceivably, the persistence after renormalization of this divergence is due to an unnatural use at small scales of an effective model meant to describe sensible physics only at large ones (cf. Ref.~\cite{Scar}).

\subsection*{Acknowledgment}
\vspace{-0.1cm}
Work supported by the National Group of Mathematical Physics (GNFM-INdAM).
\vspace{0.3cm}



\begin{thebibliography}{1}


\bibitem{AlbHeat} S. Albeverio, Z. Brze\'{z}niak, L. Dabrowski, \textsl{Fundamental solution of the heat and Schr\"odinger equations with point interaction}, J. Funct. Anal. \textbf{130}(1), 220--254 (1995).

\bibitem{Alb2} S. Albeverio, C. Cacciapuoti, M. Spreafico, \textsl{Relative partition function of Coulomb plus
delta interaction}, pp. 1--29 in J. Dittrich, H. Kova\v{r}\'{i}k, A. Laptec (Eds.), ``Functional Analysis and Operator Theory for Quantum Physics. A Festschrift in Honor of Pavel Exner'', Eur. Math. Soc. Publishing House, Z\"urich (2016).

\bibitem{Alb1} S. Albeverio, G. Cognola, M. Spreafico, S. Zerbini, \textsl{Singular perturbations with boundary conditions and the Casimir effect in the half space}, J. Math. Phys. \textbf{51}(6), 063502 (2010).

\bibitem{AlbB} S. Albeverio, F. Gesztesy, R. H{\o}egh-Krohn, H. Holden, ``Solvable Models in Quantum Mechanics'', Springer-Verlag, New York (1988).

\bibitem{BeFa} F. A. Berezin, L. D. Faddeev, \textsl{A remark on Schrodinger's equation with a singular potential}, Dokl. Akad. Nauk Ser. Fiz. \textbf{137}, 1011--1014 (1961); translation of Sov. Math. Dokl. \textbf{2}, 372--375 (1961).

\bibitem{BM2015} M. Bordag, J.  M. Mu\~{n}oz-Casta\~{n}eda, \textsl{Dirac lattices, zero-range potentials, and self-adjoint extension}, Phys. Rev. D {\bf 91}(6), 065027 (2015).

\bibitem{BorPir} M. Bordag, I.  G. Pirozhenko, \textsl{Casimir effect for Dirac lattices}, Phys. Rev. D {\bf 95}(5), 056017 (2017).

\bibitem{CFP} C. Cacciapuoti, D. Fermi, A. Posilicano, \textsl{Relative-zeta and Casimir energy for a semitrasparent hyperplane selecting transverse modes}, pp. 71--97  in G. F. Dell’Antonio, A. Michelangeli (Eds.), ``Advances in Quantum Mechanics: Contemporary Trends and Open Problems'', Springer, New York (2017).

\bibitem{FP1} D. Fermi, L. Pizzocchero, \textsl{Local zeta regularization and the Casimir effect}, Prog. Theor. Phys. \textbf{126}(3), 419--434 (2011).

\bibitem{FP2} D. Fermi, L. Pizzocchero, \textsl{Local zeta regularization and the scalar Casimir effect III. The case with a background harmonic potential}, Int. J. Mod. Phys. A \textbf{30}(35), 1550213 (2015).

\bibitem{FP3} D. Fermi, L. Pizzocchero, \textsl{Local zeta regularization and the scalar Casimir effect IV. The case of a rectangular box}, Int. J. Mod. Phys. A \textbf{31}(4\&5), 1650003 (2016).

\bibitem{FPBook} D. Fermi, L. Pizzocchero, ``Local Zeta Regularization and the Scalar Casimir Effect. A General Approach Based on Integral Kernels'', World Scientific Publishing Co., Singapore (2017).

\bibitem{FPDelta} D. Fermi, L. Pizzocchero, \textsl{Local Casimir effect for a scalar field in presence of a point impurity}, Symmetry {\bf 2018},\! {\it 10}(2), 38 (2018).

\bibitem{Grats1} Yu. V. Grats, \textsl{Casimir energy in contact-interaction models}, Phys. Atom. Nucl. {\bf 81}(2), 253--256 (2018).

\bibitem{Grats2} Yu. V. Grats, \textsl{Vacuum polarization in a zero-range potential field}, Phys. Atom. Nucl. {\bf 82}(2), 153--157 (2019).

\bibitem{Scar} A. Scardicchio, \textsl{Casimir dynamics: interactions of surfaces with codimension $>1$ due to quantum fluctuations}, Phys. Rev. D {\bf 72}(6), 065004 (2005).

\bibitem{SprZer} M. Spreafico, S. Zerbini, \textsl{Finite temperature quantum field theory on noncompact domains and
application to delta interactions}, Rep. Math. Phys. \textbf{63}(1), 163--177 (2009).

\end{thebibliography}
\end{document}